\documentstyle[aps]{revtex}
\input{epsfig}

\begin{document}

\textheight 8.8in
\textwidth 6.5in
\topmargin -.25in
\oddsidemargin -.25in
\evensidemargin 0in
\baselineskip 14pt
\def\hm{\ \rm {\it h}^{-1} Mpc}
\newcommand{\ltsima} {$\; \buildrel < \over \sim \;$}
\newcommand{\gtsima} {$\; \buildrel > \over \sim \;$}
\newcommand{\lta} {\lower.5ex\hbox{\ltsima}}
\newcommand{\gta} {\lower.5ex\hbox{\gtsima}}

\title{On the dark energy clustering properties}

\author{Francesca Perrotta$^{1,2}$\footnote{perrotta@sissa.it}
\& Carlo Baccigalupi$^{2}$\footnote{bacci@sissa.it}} 
\address{$^{1}$ INAF-Osservatorio Astronomico di Padova, 
Vicolo dell'Osservatorio 5, I-35122 Padova, Italy;\\
$^{2}$ SISSA/ISAS, Via Beirut 4, 34014 Trieste, Italy}
\baselineskip 10pt
\maketitle

\begin{abstract}
We highlight a viable mechanism leading to 
the formation of dark energy structures on sub-horizon cosmological 
scales, starting from linear perturbations in scalar-tensor cosmologies. 
We show that the coupling of the dark energy scalar field, or 
Quintessence, to the Ricci scalar induces a ``dragging" of its 
density perturbations through the general relativistic gravitational 
potentials. We discuss, in particular, how this process forces dark 
energy to behave as a pressureless component if the cosmic 
evolution is dominated by non-relativistic matter. 

This property is also analyzed in terms of the effective sound 
speed of the dark energy, which correspondingly approaches the 
behavior of the dominant cosmological component, being 
effectively vanishing after matter-radiation equality. 

To illustrate this effect, we consider Extended 
Quintessence scenarios involving a quadratic coupling between 
the field and the Ricci scalar. 
We show that Quintessence density perturbations 
reach non-linearity at scales and redshifts 
relevant for the structure formation process, respecting 
all the existing constraints on scalar-tensor theories of 
Gravity. 

This study opens new perspectives on the standard picture of 
structure formation in dark energy cosmologies, since the Quintessence 
field itself, if non-minimally coupled to Gravity, may undergo 
clustering processes, eventually forming density perturbations exiting 
from the linear regime. A non-linear approach is then required 
to further investigate the evolution of these structures, and 
in particular their role in the dark haloes surrounding galaxies 
and clusters. 
\end{abstract}

\section{Introduction}

The role of vacuum energy in cosmology is receiving a renewed 
interest since at least three cosmological observables 
indicated that a relevant fraction of the energy 
density in the Universe is presently in the form of a sort of vacuum 
energy, which is commonly known as dark energy. Type Ia Supernovae 
observations suggest that cosmic expansion is accelerated \cite{PERL,RIESS}; 
recently it has been also noticed that data indicate that acceleration is a 
relatively recent occurrence in the cosmological evolution \cite{TURNER}.
Moreover, present data on Cosmic Microwave Background (CMB) 
anisotropies favor a total energy density which is very close 
to the critical value and which is made at roughly $70\%$ by
a vacuum energy component \cite{CMBDATA}. Finally, large scale structure 
observations suggest a universe with a sub-critical matter density \cite{LSS}. 

These evidences represent a great stimulus for theoretical work, 
because understanding why the vacuum energy is at the level of the 
critical density today or less represents one of the 
main mysteries in modern fundamental physics \cite{CCP}. 
At a quantum level, a ``fine-tuning" mechanism is needed to explain 
why vacuum energy is so low with respect to any natural scale 
of vacuum expectations of fundamental fields: this discrepancy 
reaches 120 orders of magnitude if the present cosmological critical 
density is compared with the Planck scale. 
Moreover, if the mentioned cosmological observables are interpreted 
correctly, it is necessary to justify the ``coincidence" with 
which vacuum energy is starting to dominate the cosmic expansion right 
now. Describing dark energy as a scalar field $\phi$, or Quintessence, 
first considered in \cite{RP,W}, has 
the attractive feature to alleviate these problems, at least at 
classical level; ``tracking" \cite{RP,TRACK} and ``scaling" 
\cite{W,LS} solutions existing for Quintessence with inverse power 
law and exponential potentials respectively, allow dark energy 
density to be at the level of other cosmological components in the 
very early Universe. $k$-Essence models involving a generalized 
form of the kinetic energy of the scalar field can provide a mechanism
to justify the present level of dark energy \cite{K0,K}. 

Dark energy cosmologies have been considered in the  
general framework of scalar-tensor theories of Gravity. These
``Extended Quintessence" scenarios \cite{EQ} (hereafter EQ, while 
we shall refer to minimally coupled Quintessence as Q models), 
in which the scalar field responsible for cosmic acceleration 
possesses an explicit coupling with the Ricci scalar, have been 
studied under several perspectives 
\cite{UZANNMC,AMENMC,CHIBANMC,BP,RU,CHIBACOS}, 
including a detailed study of tracking trajectories 
and of their effects on cosmological perturbations \cite{TEQ}; 
in particular, both for Q and for EQ models, characteristic 
signatures have been accurately predicted on the CMB spectrum 
of anisotropies, and compared with existing data, 
see \cite{BCMB} and references therein. 

Elevating the cosmological vacuum energy to a dynamical role 
through its representation as a scalar field introduces 
the natural question about its relation with the other 
main dark component which is currently under study in 
cosmology, i.e. the Cold Dark Matter (CDM). 
The latter is currently thought to be
made of non-relativistic particles, possibly generated during the process of 
supersymmetry breaking in the early Universe, see e.g. \cite{CDM}, 
which are supposed to form the
halos around galaxies via gravitational collapse of primordial
linear perturbations. 
It is natural to ask if the formation of matter clumps in the Universe
can have some effect on dark energy, or even if dark energy clumps 
can form starting from primordial linear perturbations. 
In the past this intriguing issue 
has been faced under different points of view, either by
investigating the structure and stability of non-linear spherically 
symmetric scalar field overdensities \cite{WETTECLUMPS}, or 
by describing dark matter and dark energy as two different 
smoothly distributed classical fields \cite{MATOS,MATOS2}; 
in \cite{ALS}, clumps of an extremely light classical scalar field have 
been proposed as a candidate for Dark matter in galactic halos. 
In \cite{LH}, the formation of matter structures has been 
faced in a background filled by matter and a homogeneous 
Quintessence, which still doesn't undergo structure formation 
processes. 

In any case, the occurrence of eventual dark energy clumps arising from 
perturbations in linear regime did not 
receive any explanation. Even more, in Ref.\cite{GDM}, an effective fluid 
with negative
equation of state has been considered as 
``generalized dark matter" in the general context
of cosmological linear perturbation theory, and the main
consequences for structure formation theory have been
obtained. In particular, it has been shown that perturbations of 
a minimally coupled scalar field playing the role of 
Quintessence behave as scalar radiation on sub-horizon 
cosmological scales, relativistically dissipating scalar field density 
fluctuations, so that ordinary Quintessence rapidly becomes a smooth 
component. In the present work we investigate the sub-horizon 
dark energy perturbation behavior in scalar-tensor cosmologies, and we 
show that the conclusions can be much different. In particular, our aim 
is to give an answer to the following question: in which conditions 
it is possible to form growing dark energy density perturbations on 
sub-horizon scales, in a reference-frame independent manner? 
We consider this problem in the general context of linear perturbation 
theory in scalar-tensor cosmology. We study the perturbation 
properties of the non-minimally coupled dark energy field, 
focusing on the influence that metric 
fluctuations can have on its sub-horizon behavior; we 
estabilish in particular in which conditions such influence is 
effectively dominant, resulting in a ``gravitational dragging" of the 
dark energy itself. 
Moreover, we give a worked example of this phenomenology 
considering a typiecal Extended Quintessence model 
and giving numerical results on the sub-horizon behavior of 
its energy density fluctuations at redshifts relevant for 
structure formation. 

The results presented here are complementary with respect to our 
previous works on the same topic \cite{EQ.TEQ}: indeed, in \cite{EQ}
we wrote the basic perturbations equations, analysing the main 
cosmological features of EQ models. In \cite{TEQ}, we focused on the 
impact of tracking EQ trajectories on the CMB anisotropies and on the 
matter power spectrum. Here we deal with the clustering properties of 
the Dark energy itself, therefore completing the picture of linear 
perturbation theory in EQ cosmologies; this will require working in a 
setting in which the stress-energy tensor of the Quintessence field is 
conserved, as opposite to the formalism adopted in the previous works.  

The paper is organized as follows. In Section \ref{extended}
we review the general formalism for scalar-tensor cosmologies. 
In Section \ref{drag} we study the motion equations, both for background 
and perturbations, giving emphasis to the role of the non-minimal coupling 
in the dynamics of the scalar field density fluctuations. 
The resulting dark energy clustering properties are illustrated 
in Section \ref{secdeltarho}, where we expose some examples of these 
effects as the result of numerical integrations in typical EQ 
cosmologies. In  Section \ref{seccs} we give an equivalent interpretation 
of these results in terms of the effective dark energy sound speed. 
Finally, Section \ref{conclusions} contains the concluding remarks. 

\section{Scalar-tensor cosmologies}
\label{extended}
In this Section we give general definitions and formalism for 
describing general scalar-tensor cosmologies, both for background and 
linear perturbations. We follow, as much as possible, the notation 
adopted in original works \cite{KS,HW}. 

Scalar-tensor theories of gravity 
are generally represented by the action 
\begin{equation}
\label{action}
S=\int d^4 x \sqrt{-g} \left[ {1 \over 2\kappa} f(\phi, R) -
{\omega (\phi )\over 2} \nabla^{\mu}
\phi \nabla_{\mu}\phi -V( \phi)
+ L_{fluid}\right]\ ,
\end{equation}
where $R$ is the Ricci scalar and $\phi$ is a scalar field which 
is supposed to be coupled only with Gravity through the 
function $f(\phi ,R)$, while the functions $\omega (\phi)$ and 
$V(\phi )$ specify the kinetic and potential scalar field energies, 
respectively; the Lagrangian $L_{fluid}$ includes 
all the components but $\phi$, and the constant $\kappa$ 
plays the role of the ``bare" gravitational constant $G_{*}$, which 
in scalar-tensor theories can differ from the Newton's constant 
$G$ as it is measured by Cavendish-type experiments \cite{EFP}; 
without loss of generality, we choose the relation between $\kappa$ 
and $G_{*}$ defined in \cite{EFP} to be $\kappa =8\pi G_{*}$. 
We also pose the light velocity $c$ equal to 1. 
Einstein equations from the general action (\ref{action}) 
can be written in the familiar form 
\begin{equation}  
\label{Einstein} 
G_{\mu \nu}= \kappa T_{\mu\nu}^{total} 
\end{equation}  
with the stress-energy tensor $T_{\mu\nu}^{total}$ being made 
of the scalar field and the other components, indicated with 
the subscript {\it fluid}: 
\begin{equation} 
\label{Ttotal} 
T_{\mu\nu}^{total}=T_{\mu \nu}^{fluid}+  
T_{\mu\nu}[\phi]\ .
\end{equation}
As a consequence of the contracted Bianchi identities $T_{\mu\nu}^{total}$ 
is conserved; moreover, $T_{\mu\nu}^{fluid}$ and $T_{\mu \nu}[\phi]$ are 
separately conserved since no explicit coupling is assumed between fluid 
and $\phi$: 
\begin{equation}  
\label{conservation}
\nabla^{\mu}T_{\mu\nu}^{fluid}=\nabla^{\mu}T_{\mu\nu}[\phi ]=0\ .
\end{equation}
By defining 
\begin{equation}
\label{F}
F={1\over\kappa}{\partial f\over\partial R}\ ,
\end{equation}
the conserved scalar field contribution assumes the form 
\begin{equation} 
\label{Tphi} 
T_{\mu \nu}[\phi] = \omega\left[\nabla_{\mu}\phi \nabla_{\nu} \phi - 
{1 \over 2 } g_{\mu \nu}\nabla^{\lambda}\phi\nabla_{\lambda}\phi\right]-
Vg_{\mu \nu}+{f/\kappa -RF\over 2}g_{\mu\nu}+
\nabla_{\mu}\nabla_{\nu}F-g_{\mu \nu} \Box F + 
\left( {1\over \kappa}-F\right) G_{\mu \nu} \ ,
\end{equation} 
where we can recognize the origin of the different terms composing  
the scalar field stress-energy tensor (\ref{Tphi}): the minimal-coupling  
(including a  ``kinetic'' and a ``potential'' part), the  non-minimal  
coupling including $f$, $F$ and $R$, and the gravitational term, 
proportional to $(\kappa^{-1}-F)$, contaning the Einstein tensor itself. 
We can define them as 
\begin{equation}
\label{Tmunumc}
T_{\mu \nu}^{mc}[\phi] = \omega[\nabla_{\mu}\phi \nabla_{\nu} \phi - 
{1 \over 2 } g_{\mu \nu}\nabla^{\lambda}\phi\nabla_{\lambda}\phi]-Vg_{\mu 
\nu}\ ,
\end{equation}
\begin{equation} 
\label{Tmununmc}
T_{\mu \nu}^{nmc}[\phi]={f/\kappa -RF\over 2}g_{\mu\nu}+
\nabla_{\mu}\nabla_{\nu}F -g_{\mu \nu} \Box F\ ,
\end{equation}
\begin{equation}
\label{Tmunugrav}
T_{\mu \nu}^{grav}[\phi] =\left( {1\over \kappa}-F\right) G_{\mu \nu}\ . 
\end{equation}
It is relevant to note that, as extensively discussed in \cite{Faraoni},  
the gravitational term may also be taken to the left hand side of eq.  
(\ref{Einstein}): 
\begin{equation} 
\label{2approach} 
FG_{\mu \nu}= {\tilde{T}}_{\mu \nu}^{total} \equiv T_{\mu  
\nu}^{fluid}+T^{mc}[\phi]+T^{nmc}[\phi] 	 
\end{equation} 
With such approach, one would be left with a not 
conserved total stress-energy tensor ${\tilde{T}}_{\mu \nu}^{total}$,  
which would differ from (\ref{Ttotal}) because the absence of the merely  
``gravitational'' term in (\ref{Tphi}). Including the gravitational sector 
in the stress-energy tensor of the scalar field is not only a dichotomy, 
as we will discuss in this paper. First of all, in typical non-minimal 
coupling models in which $F$ is $\kappa^{-1}$ plus a function depending 
explicitely on $\phi$, the latter function describes 
the energy transfer between field and metric, 
acting in particular at a quantum level, see e.g. \cite{BD}. 
Second, the Bianchi identities allow us to write down conserved quantities 
for the scalar field. Third, the linear perturbation theory describes the 
evolution of small perturbation in the scalar field energy density: if the 
latter is drawn from a conserved stress-energy tensor, perturbations in 
the scalar field energy density exhibit a behavior which is hidden with 
the approach (\ref{2approach}). 

Assuming a Friedmann Robertson Walker (FRW) background, the metric 
tensor assumes the form 
\begin{equation}
\label{frw}
g_{\mu\nu}=a^{2}(\eta )(\gamma_{\mu\nu}+h_{\mu\nu})\ ,
\end{equation}
where $\eta$ is the conformal time, $a(\eta)$ is the scale factor with 
conformal time derivative $\dot{a}$ expressed through the Hubble parameter 
${\cal H}=\dot{a}/a$ and 
$\gamma_{\mu\nu}=diag[-1,(1-Kr^{2})^{-1},r^{2},r^{2}\sin^{2}\theta]$ 
is the background metric with spatial curvature $K$ in spherical 
coordinates ($\eta ,r,\theta,\phi$); $h_{\mu\nu}\ll 1$ represents linear metric 
cosmological perturbations, and is conveniently 
expressed in the Fourier space: for scalar perturbations, 
indicating with $Y$ the solution of the Laplace equation 
$\nabla^{s\, i}\nabla^{s}_{i}Y=-|\vec{k}|^{2}Y$, where 
$\nabla^{s}$ means covariant derivative with respect to 
the spatial metric $\gamma_{ij}$, the amplitude at wave vector $\vec{k}$ 
of the most general scalar metric perturbation, see e.g. \cite{KS}, can be 
written as 
\begin{equation}
\label{hmunu}
h_{00}=-2AY\ ,\ h_{0j}=-BY_{j}\ ,\ h_{ij}=2H_{L}Y\gamma_{ij}+2H_{T}Y_{ij}\ ,
\end{equation}
where $A,B,H_{L}$ and $H_{T}$ represent the amplitude in the Fourier space at 
wave vector $\vec{k}$; $Y_{j}$ and the traceless $Y_{ij}$, with $i,j=1,2,3$, 
are defined as $Y_{j}=-(1/k)\nabla^{s}_{j}Y$, 
$Y_{ij}=(1/k^{2})\nabla^{s}_{i}\nabla^{s}_{j}Y+\gamma_{ij}Y/3$. 
Note that, as customary, we 
intentionally do not write the argument $\vec{k}$ explicitly in the amplitude
of the  perturbation quantities in the Fourier space. As it is well known
\cite{KS}, a gauge freedom exists because of the linearization of the
problem, so that two of the four quantities in (\ref{hmunu}) can be set to 
zero, or, equivalently, two independent gauge invariant combinations 
can be built out of $A,B,H_{L}$ and $H_{T}$. Correspondingly, the stress energy 
tensor $T_{\mu\, x}^{\nu}$ relative to any cosmological component $x$ splits in a
background component  $T_{\mu\, x}^{\nu}=diag[-\rho_{x},p_{x},p_{x},p_{x}]$ 
and perturbations $\delta T_{\mu\, x}^{\nu}$  represented as 
\begin{equation}
\label{deltaTmunu}
\delta T_{0\, x}^{0}=-\rho_{x}\,\delta_{x}\,Y
\ ,\ \delta T_{j\, x}^{0}=(\rho_{x}+p_{x})(v_{x}-B)Y_{j}
\ ,\ \delta T_{j\, x}^{i}=
p_{x}(\pi_{L\, x}\delta_{j}^{i}+\pi_{T\, x}Y_{j}^{i})\ ,
\end{equation}
where in particular $\delta_{x}$ represent the density contrast fluctuation 
at wave vector $\vec{k}$. It is also useful to introduce the fluctuations 
in the expectation value of the scalar field at wave vector $\vec{k}$, 
which will be indicated as $\delta\phi\, Y$. 

Einstein and conservation equations (\ref{Einstein},\ref{conservation}) 
split into two separate sets describing the evolution of background 
and perturbations. In the next Section we'll write them explicitly, 
focusing on the role of the different quantities composing the 
scalar field stress energy tensor $T_{\mu\nu}[\phi ]$ defined in (\ref{Tphi}). 

\section{Gravitational dragging}
\label{drag}

Let us then concentrate on the conserved tensor (\ref{Tphi}), first 
considering background quantities and their evolution equations. 
Conservation laws (\ref{conservation}) for the unperturbed 
scalar field reduce to 
\begin{equation}  
\label{continuitybackground} 
\dot{\rho}_{\phi}=-3{\cal H}(1+w_{\phi})\rho_{\phi}\ , 
\end{equation} 
having defined the dark energy field equation of state as 
$w_{\phi}=p_{\phi}/ \rho_{\phi}$. 
The background evolution equations will be completely determined 
by the Friedmann-Robertson-Walker equations 
\begin{equation} 
\label{FRWequations}
{\cal{H}}^2= {a^{2}\kappa \over 3}\left(\rho_{fluid}+\rho_{\phi}\right)+K
\ ,\ \dot{\cal{H}}= -{a^{2}\kappa \over 6}
\left(\rho_{fluid}+\rho_{\phi}+3p_{fluid}+3p_{\phi}\right)\ ,
\end{equation} 
and by the conservation equations for each component $x$: 
$\dot{\rho}_{x}=-3{\cal{H}}({\rho}_x+p_x )$.
As it can be easily seen from (\ref{Tphi}), the expressions for the scalar 
field energy density and pressure which satisfy (\ref{continuitybackground}) 
are given by  
\begin{equation} 
\label{rhophi}
\rho_{\phi}=\omega {{\dot{\phi}}^2 \over 2 a^2}+V(\phi)+
{RF-f/\kappa\over 2a^{2}}-
{3\over a^{2}}{\cal H}\dot{F}+3{{\cal H}^{2}+K\over a^{2}}
\left( {1\over\kappa}-F \right)\ ,
\end{equation}
\begin{equation}
\label{pphi} 
p_{\phi}=\omega {\dot{\phi}^2 \over 2 a^2}-V(\phi)-
{RF-f/\kappa\over 2a^{2}}+{1\over a^{2}}\left({\cal H}\dot{F}+\ddot{F}\right)-
{2\dot{\cal H}+{\cal H}^{2}+K\over a^{2}}\left({1\over\kappa} -F\right)\ .
\end{equation}
It is useful to mention that these expressions combine in the continuity 
equation (\ref{continuitybackground}) to give the Klein-Gordon equation: 
\begin{equation}
\label{KG}
\ddot{\phi}+2 {\cal{H}} \dot{\phi}+{1 \over 2 \omega}
\left(\omega_{,\phi}\dot{\phi}^2 -{a^2\over\kappa}f_{,\phi}+
2a^2 V_{,\phi}\right)=0\ .
\end{equation}
Before considering perturbed quantities, it is relevant to comment 
briefly on the role of the gravitational component 
of the scalar field stress-energy tensor (\ref{Tmunugrav}), since 
as we'll see in a moment the same arguments hold for perturbations. 
As first noted in \cite{RU,CHIBACOS}, under conditions in which $F$ differs 
from $\kappa^{-1}$, even by a small amount due to a non-zero value of $\phi$, 
the gravitational term appearing in the expression (\ref{rhophi}) switches on, 
feeding the scalar field energy density itself with a term 
proportional to the square of the real time Hubble parameter 
$H={\cal H}/a$, which in turn is proportional to the {\it total} 
cosmological energy density through the Einstein equations. Since the latter 
is made of matter and radiation scaling as $1/a^{3}$ and
$1/a^{4}$ respectively, it is straightforward that at sufficiently early times 
the gravitational term dominates the dynamics of $\rho_{\phi}$. 
As we'll see, this process, which can be meaningfully named 
``gravitational dragging", is also very important for the dynamics 
of the dark energy perturbations. 

Equations $\nabla_{\mu}\delta T_{0}^{\mu}[\phi ]=0$ and 
$\nabla_{\mu}\delta T_{j}^{\mu}[\phi ]=0$ correspond respectively 
to the continuity and Euler equations 
\begin{equation}
\label{continuity}
\left({\delta_{\phi}\over 1+w_{\phi}}\right)^{\cdot}+kv_{\phi}+
3\dot{H}_{L}+3{\cal H}{w_{\phi}\over 1+w_{\phi}}\Gamma_{\phi}=0\ ,
\end{equation}
\begin{equation}
\label{euler}
(\dot{v}_{\phi}-\dot{B})+
{\cal H}\left(v_{\phi}-B\right)\left(1-3c_{s\,\phi}^{2}\right)-
k{1\over 1+w_{\phi}}\pi_{L\,\phi}+
{2\over 3}{1\over 1+w_{\phi}}{k^{2}-3K\over k}\pi_{T\,\phi}-kA=0\ ,
\end{equation}
where we have defined the scalar field entropy perturbation 
\begin{equation}
\label{entropy}
\Gamma_{\phi}=\pi_{L\,\phi}-{c_{s\,\phi}^{2}\over w_{\phi}}\delta_{\phi}\ ,
\end{equation}
and its sound velocity 
\begin{equation}
\label{cs2}
c_{s\,\phi}^{2}={\dot{p}_{\phi}\over\dot{\rho}_{\phi}}\ .
\end{equation}
Equations (\ref{continuity},\ref{euler}) hold formally for any cosmological 
component. As for the background, they combine in the perturbed 
Klein-Gordon equation 
$$
\ddot{\delta\phi}+\left(2{\cal H}+
{\omega_{,\phi}\over\omega}\dot{\phi}\right)\dot{\delta\phi}+ 
\left[k^{2}+
\left({\omega_{,\phi}\over\omega}\right)_{,\phi}{\dot{\phi}^{2}\over 2}+
\left({-a^{2}f_{,\phi}/\kappa+2a^{2}V_{,\phi}\over 2\omega}\right)_{,\phi}
\right]\delta\phi =
$$
\begin{equation}
\label{perturbedKG}
=\dot{\phi}\dot{A}-
\left(3{\cal H}\dot{\phi}+
{-a^{2}f_{,\phi}/\kappa+2a^{2}V_{,\phi}\over 2\omega}
\right)A+\dot{\phi}\left(3{\cal H}A-3\dot{H}_{L}-kB\right)+
{1\over 2\omega\kappa}f_{,\phi R}\delta R\ ,
\end{equation}
with the variation of the Ricci scalar given by 
$$
\delta R=-{2\over a^{2}}\left(3{\cal 
H}A-3\dot{H}_{L}-kB\right)^{\cdot}Y-
{6\over a^{2}}{\cal H}\left(3{\cal H}A-3\dot{H}_{L}-kB\right)Y+
$$
\begin{equation}
\label{deltaR}
+{2\over a^{2}}\left(k^{2}-3\dot{\cal H}+3{\cal H}^{2}\right)AY+
{4\over a^{2}}\left(k^{2}-3K\right)\left(H_{L}+{1\over 3}H_{T}\right)Y\ .
\end{equation}
In order to gain insight into the behavior of the scalar field 
perturbations specifically, let us write explicitly the formal solutions 
to the above equations. The variation 
of the stress-energy tensor (\ref{Tphi}) yields contributions 
which we classify in {\it mc}, {\it nmc} and {\it grav} as we did 
in equations (\ref{Tmunumc},\ref{Tmununmc},\ref{Tmunugrav}). 
Let us start from the energy density perturbations: 
$\delta T^0_0[\phi]=-\rho_{\phi}\delta_{\phi}Y=
\delta T_0^{0\, mc}[\phi ]+\delta T_0^{0\, nmc}[\phi ]+
\delta T_0^{0\, grav}[\phi ]$; the different contributions 
are given by 
\begin{equation}
\label{dT00mc}
\delta T_0^{0\, mc}[\phi]=\left[-\omega_{,\phi} {\dot{\phi}^2 \over 2 a^2} 
\delta  \phi+{\omega \over a^2}\left( A \dot {\phi}^2-\dot{\phi}\delta 
\dot{\phi}\right)-V_{,\phi}  \delta  \phi  \right] Y
\end{equation} 
\begin{equation}
\label{dT00nmc}
\delta T_0^{0\, nmc}[\phi]=\left[-{3\over a^2}A{\cal H}\dot{F}+
{3\over a^{2}}{\cal{H}}\delta \dot{F} + {1\over 2\kappa}f_{,\phi}\delta\phi+
\left(-{R\over 2}+{k^2 \over a^2 }\right) \delta F+
{3\over a^{2}} {\cal H} {\cal{K}}_g \dot{F} \right] Y
\end{equation} 
\begin{equation}
\label{dT00grav}
\delta T_0^{0\, grav}[\phi]={3\over a^{2}}\left({\cal H}^2+K\right)\delta F Y 
+\left({1\over \kappa}-F\right){2 \over a^2} \left[ 3 {\cal{H}}^2 A - {\cal{H}}
kB -3 {\cal{H}} \dot{H}_L -\left(k^{2}-3K\right)
\left( H_L+{H_T \over 3} \right)\right]Y  \end{equation} 
where we have defined ${\cal{K}}_g \equiv -A +{\cal{H}}^{-1}B/3 + 
{\cal{H}}^{-1}\dot{H}_L$; note also that in general 
$\delta F=F_{,\phi}\delta\phi +F_{,R}\delta R$. 
The momentum perturbation $\delta T^{0}_{j}[\phi ]$ is composed by 
\begin{equation} 
\label{dt0jmc}
\delta T^{0\, mc }_{j} = {k \over a^2 }
\left(\omega \dot{\phi}\delta\phi\right)Y_{j}\ ,
\end{equation} 
\begin{equation} 
\label{dT0jnmc}
\delta T^{0\, nmc}_{j} = {k \over a^2 }\left(\delta\dot{F}-
{\cal H}\delta F-A\dot{F}\right)Y_{j}\ ,
\end{equation} 
\begin{equation} 
\label{dT0jgrav}
\delta T^{0\, grav}_{j} = {2\over a^2}
\left( {1\over \kappa}-F\right) \left(k{\cal{H}}A- 
k\dot{H}_L-\left({k^{2}-3K\over 3k}\right)
\dot{H}_T\right)Y_{j}\ .
\end{equation} 
Let us consider now $\delta T^{i}_{j}$. In general, it will have both trace 
and traceless components, $\pi_{L\,\phi}$ and $\pi_{T\,\phi}$ respectively, 
as in equation (\ref{deltaTmunu}). 
The different contributions are given by 
\begin{equation}
\label{dTijmc}
\delta T^{i\, mc}_j [\phi]= {1\over a^{2}}
\left[ \omega  \dot{\phi} \delta \dot{\phi} + 
{\omega_{,\phi} \over 2} 
\dot{\phi}^2\delta\phi -a^2 V_{,\phi} \delta \phi -A\omega\dot{\phi}^{2}
\right] Y \delta^i_j \ ,
\end{equation}
$$
\delta T^{i\, nmc}_j [\phi]={1\over a^{2}}
\left[{1\over 2\kappa}f_{,\phi}\delta\phi +
2\dot{F}{\cal{H}}{\cal{K}}_g+\left({R\over 2}+
{2\over 3} k^2\right) \delta F + {\cal H} \delta
\dot{F}+\delta 
 \ddot{F}-\dot{F} \dot{A}-2 \ddot{F} A  \right] Y \delta^i_j+
$$
\begin{equation} 
\label{dTijnmc}
+{k^{2}\over a^{2}}
\left[ \delta F +\left(kB-\dot{H}_T\right)
{\dot{F} \over k^2}\right]
Y^i_j\ ,
\end{equation}
$$
\delta T^{i\, grav}_j [\phi]={1\over a^{2}}
\left(2\dot{{\cal{H}}} +{\cal{H}}^2+2K\right)
\delta FY\delta^i_j +
$$
$$
+{2\over a^{2}}\left({1\over\kappa}-F\right)
 \left[ \left(2\dot{{\cal{H}}} +{\cal{H}}^2\right) A 
+{\cal{H}}\dot{A}-{k^2  \over 3} A-{k \over 3}\dot{B} -
{2 \over 3}k  {\cal{H}}B -
\ddot{H}_L-2{\cal{H}}\dot{H}_L- 
{k^2 -3K\over  3}  \left( H_L+{H_T \over 3}\right)
 \right]Y\delta^i_j+
$$
\begin{equation}
\label{dTijgrav}
+{1\over a^{2}}\left({1\over \kappa}-F\right)\left[-k^2 A-k\left( 
\dot{B}+{\cal{H}}B \right)+\ddot{H}_{T}+
{\cal{H}}(2\dot{H}_T-kB)-k^2 \left(H_L+{H_T \over 3} \right)
\right] Y^i_j\ ,
\end{equation}
where $\pi_{L\,\phi}$ and $\pi_{T\,\phi}$ are given by the terms 
proportional to $\delta_{j}^{i}$ and $Y_{j}^{i}$, respectively. 
It is worth to note that an interesting feature of non-minimally 
coupled scalar fields is the presence of the gauge invariant anisotropic 
stress $\pi_T$: as shown in \cite{GDM} and \cite{GDM2}, stress 
perturbations have a role in the structure formation; we will return 
to this in Sec.\ref{seccs}. 

Although the expressions above appear complicated, it is quite simple 
to highlight the point we are interested in. In the quantities 
(\ref{dT00grav},\ref{dT0jgrav},\ref{dTijgrav}), the terms multiplying 
$(1/\kappa -F)$ are $\delta G_{0}^{0},\delta G^{0}_{j},\delta G^{i}_{j}$, 
respectively. Focusing on the gravitational part of the scalar field 
energy density perturbation, as we noted above for the case of the 
background quantities, if $F$ differs from $1/\kappa$ the scalar field 
density perturbation is fed by the {\it total} density fluctuation, 
since $\delta G_{0}^{0}=\kappa\delta T_{0}^{0}$. Therefore, if this term 
dominates over the others ($mc,nmc$), the gravitational dragging is active on 
the density perturbations and forces the scalar field density fluctuations 
to behave as the dominant cosmological component. 

As we'll see in the next Section, this process becomes crucially important
in dark energy cosmologies, where the scalar field plays an important role 
in the cosmic evolution, determining the cosmic acceleration today. 
In the next Section we will give a worked example of this issue. 
We integrate Einstein and conservation equations to get the
cosmological evolution in a typical Extended Quintessence scenario where 
$1/\kappa -F\propto\phi^{2}$, focusing on the behavior of the dark energy 
density fluctuations at the relevant redshifts for structure 
formation. 

\begin{figure} 
\centerline{ 
\psfig{file=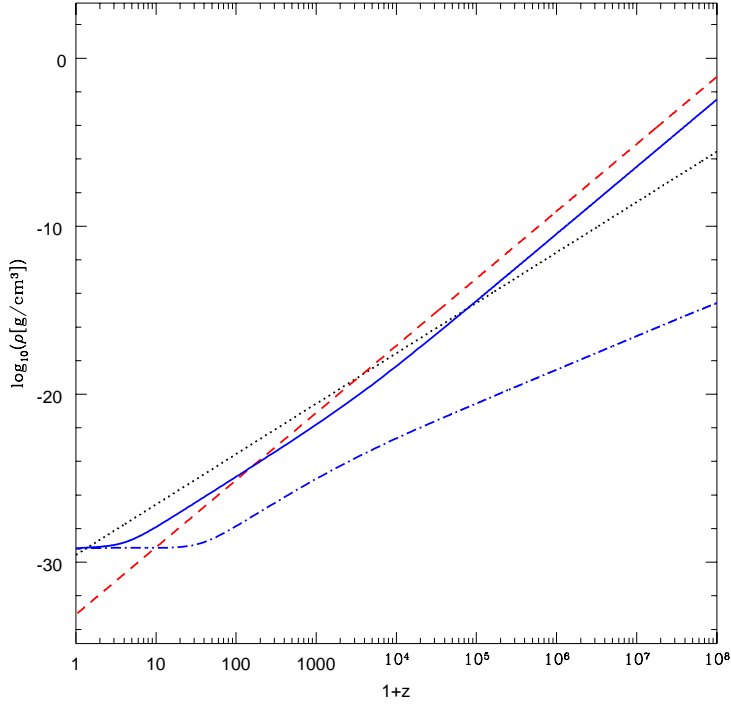,height=4.in,width=4.in} 
} 
\caption{Redshift scaling of cosmological components 
in tracking EQ scenario (see text): radiation (dashed), matter 
(dotted), total dark energy as from eq.(\ref{rhophi})(solid), kinetic 
and potential dark energy contributions (first two terms in  
eq.(\ref{rhophi})) (dotted dashed).} 
\label{f1}   
\end{figure} 

\section{Dark energy clustering}
\label{secdeltarho}
Let us focus on the non-minimally coupled scalar field fluctuations; 
combining (\ref{dT00mc},\ref{dT00nmc},\ref{dT00grav}) and the expression 
for the background energy density (\ref{rhophi}), one gets the following 
expression for the scalar field energy density fluctuation: 
\begin{equation}
\label{deltaphi}
\delta_{\phi}=\delta_{\phi}^{mc}+\delta_{\phi}^{nmc}+\delta_{\phi}^{grav}\ ,
\end{equation}
where 
\begin{equation}
\label{deltaphimc}
\delta_{\phi}^{mc}={\omega_{,\phi} \dot{\phi}^2 
\delta  \phi-2\omega( A \dot {\phi}^2-\dot{\phi}\delta  
\dot{\phi})+2a^{2}V_{,\phi}  \delta  \phi 
\over
\omega\dot{\phi}^{2}+2a^{2}V+RF-f/\kappa -6{\cal H}\dot{F}-2a^{2}
(1/\kappa -F)G_{0}^{0}}\ ,
\end{equation}
\begin{equation}
\label{deltaphinmc}
\delta_{\phi}^{nmc}=
{6A{\cal H}\dot{F}-6{\cal{H}}\delta \dot{F}-
a^{2}(f_{,\phi}/\kappa )\delta\phi -
(-a^{2}R+2k^{2}) \delta F-6{\cal H} {\cal{K}}_g \dot{F} 
\over
\omega\dot{\phi}^{2}+2a^{2}V+RF-f/\kappa -6{\cal H}\dot{F}-2a^{2}
(1/\kappa -F)G_{0}^{0}}\ ,
\end{equation}
\begin{equation}
\label{deltaphigrav}
\delta_{\phi}^{grav}={-6\delta F{\cal H}^{2}-
2a^{2}(1/\kappa -F)\delta G_{0}^{0}
\over
\omega\dot{\phi}^{2}+2a^{2}V+RF-f/\kappa -6{\cal H}\dot{F}-2a^{2}
(1/\kappa -F)G_{0}^{0}}\ .
\end{equation}
Again, we point out that this is precisely the form obtained perturbing 
the field energy density, whenever the latter is drawn from a conserved 
energy-momentum tensor: only in this case, we are allowed to use eq. 
(\ref{continuity}) for the field energy density evolution. Most importantly, 
the use of conserved quantities allows a more direct interpretation of the 
interchange between different species. The gauge invariant total 
density perturbation $\Delta$ and the gravitational potential $\Phi$, 
defined by 
\begin{equation}
\label{rhodelta}
\rho\Delta =\sum_{x}\left[\rho_{x}\delta_{x}+
3{{\cal H}\over k}\left(\rho_{x}+p_{x}\right)v_{x}\right]-
\left(\rho +p\right)B\ , 
\end{equation}
\begin{equation}
\label{Phi}
\Phi=
H_{L}+{H_{T}\over 3}+{{\cal H}\over k}\left(B-{\dot{H}_{T}\over k}\right)\ ,
\end{equation}
are related through the Einstein equation 
\cite{Bardeen,KS}: 
\begin{equation}
\label{Poisson}
\Phi={\kappa a^2 \over 2 k^2}  \rho \Delta\ .
\end{equation}
In such a way, since $\Delta$ sums up perturbations in all the fluid 
components, a ``potential well'' may be generated by each of them (in 
particular, by a perturbation in the scalar field energy density), 
affecting the behavior of density perturbations in all the other species 
(in particular, matter perturbations). Viceversa, perturbations in the 
matter component will perturb the gravitational potential to drive the 
scalar field energy density perturbations: such a kind of back-reaction is 
precisely what we expect by looking at equation (\ref{deltaphigrav}), due 
to the presence of the $\delta G^0_0$ term. When the energy density 
perturbations of the total fluid are dominated 
by perturbations in the matter component (i.e. at sufficiently high 
redshifts in typical dark energy cosmologies) for some scale $k^{-1}$, 
the term $\delta G^0_0 =\kappa\delta T^0_0$ in (\ref{deltaphi}) 
is in turn dominated by matter 
energy density perturbations, which then act as a source of the scalar 
field density perturbations. 

The very interesting feature here is that non-vanishing energy density 
perturbations of a non-minimally coupled scalar field can 
even be associated to a homogeneous scalar field as long as a non-zero 
value of $\phi$ makes $\kappa^{-1}\ne F$: we can easily see that  
$\delta\rho_{\phi}$ in equation (\ref{deltaphi}) survives even in the limit 
$\delta \phi  \rightarrow 0$, because of the gravitational dragging. 
In other words, perturbations of a non-minimally coupled 
scalar field are sourced by two complementary mechanisms: proper scalar 
field perturbations, and metric induced perturbations, related to the 
Ricci scalar coupled with the field itself. 

Focusing now on dark energy cosmologies, the described process introduces 
genuine new features with respect to ``ordinary'' (minimally coupled) 
Quintessence scenarios:  the growth in the matter perturbations may drag EQ 
density  perturbations to a non-linear regime, opening the possibility of the 
formation of Quintessence clumps. 

To give a concrete example, we numerically evolve linear perturbations 
in a typical EQ scenarios \cite{TEQ}. The coupling of Quintessence with 
the Ricci scalar is chosen to have the structure 
\begin{equation}
\label{FEQ}
{1\over\kappa}f(\phi ,R)=F(\phi )R\ .
\end{equation}
The measured gravitational constant $G$, in scalar-tensor theories 
(\ref{action})  with the choice (\ref{FEQ}), is related to the various 
quantities in the Lagrangian  as follows \cite{EFP}: 
\begin{equation}
\label{Geff}
G={G_{*}\over\kappa F}
\left( {2\omega F+4{F_{,\phi}}^2 \over 2 \omega F+3{F_{,\phi}}^2} \right)\ .
\end{equation}
As in \cite{EQ,TEQ}, we model $F$ as a constant plus a 
term yielding a quadratic coupling between the field and the Ricci scalar, 
so that 
\begin{equation}
F(\phi )={1\over\kappa}+\xi\phi^{2}\ ,
\end{equation}
where $\xi$ is the non-minimal coupling constant, with the constraint 
that today $F$ satisfy the relation (\ref{Geff}). 
Note also that $\xi$ can in principle assume 
both positive and negative values. Moreover, its magnitude is not arbitrary, 
due to constraints from local and solar-system experiments on the time-variation  
of the gravitational constant and from effects induced on photon  
trajectories \cite{JBD}. The time derivative $G_{t}$ of the observed 
gravitational constant $G$ as defined in equation (\ref{Geff}) must satisfy 
$G_{t}/G\lta 10^{-12}$ yr$^{-1}$ at present. Moreover the Jordan-Brans-Dicke 
parameter $\omega_{JBD}=\omega F/F_{,\phi}^{2}$ must be greater than about 
2500 at  present; to be conservative, we choose $\omega_{JBD}=3000$, with a 
negative  sign of the coupling constant (in our specific model this 
corresponds to $\xi\simeq -1.78\cdot 10^{-2}$, $\phi_{0}\simeq 1/\sqrt{G}$), 
which yields $G_{t}/G\simeq 10^{-14}$ yr$^{-1}$. The Quintessence potential, 
responsible for cosmic acceleration today, has an inverse power law form 
$V\propto\phi^{-\alpha}$; moreover, we fix $\omega (\phi )=1$. 

The cosmological model is specified as follows. 
The Hubble parameter at present is fixed at $H_{0}=100h$ km/sec/Mpc 
with $h=0.7$ and the spatial metric is taken to be flat, $K=0$. 
The fraction of critical density in dark energy is $\Omega_{\phi}=0.70$. 
The equation of state of Quintessence at present $w_{\phi\, 0}$ is 
chosen to be $-0.9$, yielding cosmic acceleration. Baryon abundance 
is set to $\Omega_{b}h^{2}=0.022$, CDM represents 
the remaining matter component, $\Omega_{CDM}=1-\Omega_{\phi}-\Omega_{b}$, 
and three massless neutrino families are assumed. 
Perturbations are taken to be Gaussian with an initially scale 
invariant adiabatic spectrum \cite{PB}. 
The evolution of background and perturbations has been 
determined by numerically solving equations in the synchronous gauge 
$A=B=0$. Their expressions are reported in the Appendix. 

Let us consider the background evolution first. In figure \ref{f1} the 
energy densities of radiation (dashed line), matter (dotted) and dark 
energy (solid) are plotted as a function of the redshift $z=1/a -1$. 
At late times, $z\lta 5$, the dark energy density is dominated by the 
kinetic and potential energies. 
At higher redshift the effect of the gravitational dragging is evident: 
the last term in (\ref{rhophi}) actually dominates and forces dark 
energy to scale with redshift as the dominant cosmological component. 
As it is easy to see, in this regime, the dark energy cosmological 
abundance is simply given by 
\begin{equation}
\label{omegaphiscaling}
\Omega_{\phi}(z)\simeq -\kappa\xi\phi^{2}(z)\ ,
\end{equation}
where the minus is due to our sign conventions. Note that  
the quantity $\Omega_{\phi}$ can be constrained by Big Bang 
Nucleosynthesis (BBN), because a variation in the gravitational constant 
can be regarded as inducing a change into the effective number of 
massless neutrinos \cite{NUCLEO}; however in the present case 
this value is at percent level, too small to produce observable effects. 
The dotted-dashed line represents indeed only the contribution from the 
{\it mc} terms in (\ref{rhophi}): the rising part of this curve at 
$z\gta 1000$ is due to the $R$-boost \cite{TEQ} induced by the 
effective gravitational potential in the Klein-Gordon equation (\ref{KG}). 
We stress that while the latter contribution comes from the kinetic energy 
of the field rolling on the effective gravitational potential, 
so that ultimately implies a change in time of the physical 
value of the scalar field $\phi$, the gravitational dragging can be 
thought as a power injection into the dark energy density coming 
from the total one, while it does not require directly a spatial 
dependence of the expectation value of $\phi$. Note also that a condition 
in which dark energy scales as the dominant cosmological component can be 
achieved with an exponential potential \cite{W,LS}; however, 
in that case the field is minimally coupled and Quintessence 
density perturbations vanish relativistically after horizon 
crossing \cite{GDM}. 

Let us turn now to consider the perturbations. Since we are interested in the 
dark energy clustering during the formation of matter structures, we 
concentrate on the behavior in the matter dominated era. 
Moreover, we focus on the logarithmic power of density fluctuations 
at the scale $k$, defined by  
\begin{equation}
\label{deltak}
\delta_{k\, x}^{2}=4\pi k^{3}\delta_{x}^{2}\ ,
\end{equation}
where $x$ represents a generic component. 
As it is well known, a scale for which $\delta_{k\, x}\simeq 1$ has to be 
considered in non-linear regime. In figure \ref{f2} we plot $\delta_{k}$ 
for matter (dotted) and dark energy (solid) at relevant redshifts. 
As it is expected, the gravitational dragging is active and forces 
Quintessence perturbations to behave as non-relativistic matter 
on sub-horizon scales, when the gravitational terms dominate both 
$\rho_{\phi}$ and $\delta\rho_{\phi}$. Under these conditions, 
by using (\ref{rhophi}) and (\ref{deltaphigrav}) we see that we can 
write approximately 
\begin{equation}
\label{bomba}
\delta_{\phi}\simeq\delta_m\ ,
\end{equation}
which is mostly satisfied, in our specific model, at $z\gta 5$. 
It is in fact not a case that all the quantities specifying our 
specific model disappeared in the above relation. Indeed, (\ref{bomba}) 
holds if three general conditions are satisfied in dark energy cosmologies, 
namely $(i)$ Gravity deviates from General Relativity, $(ii)$ 
Quintessence plays the role of the non-minimally coupled field and 
$(iii)$ gravitational terms dominate (\ref{deltaphi},\ref{rhophi}). 
At present, when the kinetic and potential energies of the field 
dominate the cosmic expansion imprinting acceleration, the condition 
(\ref{bomba}) is broken because matter is no more the dominant 
component. 

To fully understand the importance of this plot, we reported also 
$\delta_{k\,\phi}$ for a corresponding model in which the Quintessence 
is minimally coupled. represented by the dashed dotted curve. 
The latter is rising with time because, as noticed in \cite{BMR}, 
the inhomogeneous term of the perturbed Klein Gordon equation 
(\ref{perturbedKG}) is driving the evolution of $\delta\phi$. 
However, as it is evident from the figure, in absence of non-minimal 
coupling the dark energy density perturbations do not play any role 
in structure formation. 

Therefore, maybe the most interesting consequence of the gravitational 
dragging in dark energy cosmologies is that the non-linearity may arise for 
the Quintessence component, at a redshift depending in particular on the 
coupling strength, opening the possibility of the formation of Quintessence 
large overdensities and cavities on sub-horizon scales. 
On the other hand, at a linear level, the effect produced 
by $\delta\rho_{\phi}$ on the total gravitational potential $\Phi$ 
resides in a fraction $\rho_{\phi}/\rho_{m}$, which is 
small in the limit in which $\phi$ is subdominant. 
For example, in models in which the assumed scalar-tensor 
Gravity theory is slightly different from General Relativity, 
i.e. $|1-\kappa F|\ll 1$, in the gravitational dragging regime 
when the gravitational contribution dominate $\rho_{\phi}$ and
$\delta_{\phi}$ through the products $(\kappa^{-1}-F)G_{0}^{0}$ and 
$(\kappa^{-1}-F)\delta G_{0}^{0}$, it can be easily seen that the 
the portion of gravitational potential which is 
sourced by the Quintessence is given approximately by 
\begin{equation}
\label{Poissonphi}
\Phi_{\phi}\simeq\left(1-\kappa F\right)\Phi\ll\Phi\ ,
\end{equation}
which means that the bulk of the gravitational potential arising from 
the clustering process is still provided by matter. Note however that 
this is true for linear perturbations, which we are treating in this 
work; the effect a non-minimally coupled dark energy on the gravitational 
potential associated to a non-linear structure is still unknown, 
and actually this study could be an interesting development of the 
present work. 

The possibility of the presence of Quintessence clumps in the 
Universe has been considered by several authors, mainly aimed to foresee 
their signatures on the Galactic structure \cite{WETTECLUMPS,MATOS,MATOS2}; 
however, there is not, at the present, a theory explaining how such 
``vacuum energy clumps'' may form, and on which scales we expect eventually to
find them at low redshifts. What we obtained is just a possible way to escape
from the linear regime. In other models, as in \cite{BP}, there can be even
higher deviations from General Relativity at high redshift, than in the
model considered here; the potential effect on the formation of non-linear
clumps may be even more important. Moreover, we have shown that the occurrence 
of dark energy non-linear sub-horizon structure does not require directly a 
space dependence of the expectation value of $\phi$. 

Obviously, this is just a first step towards a theory of ``vacuum energy 
clumps'': the perturbation behavior out of the linear regime is still an 
open issue, and is strictly related to properties such as the sound speed 
of the dark energy component. In particular, we think that one of the key 
issues is the time needed for a Quintessence primordial structure to 
collapse, and a fundamental role on this is played not only by the bare 
coupling  constant $\xi$, but generally by the whole coupling function,  
through the effect they have on the value, and sign, of the 
Quintessence sound speed. In the next Section we'll use a different 
approach to explain such properties, based on the effective sound speed 
introduced in \cite{GDM,GDM2}. 

\section{Dark energy sound speed}
\label{seccs}
An important role in the perturbation growth is played 
by the non adiabatic stress or entropy contribution, entering 
directly into the evolution equation for density perturbations; 
by defining $\delta p_{x}=p_{x}\pi_{L\, x}$ for a generic 
component $x$ we can write the scalar field entropy contribution as 
\begin{equation}
\label{entropybis}
p_{\phi} \Gamma_{\phi}=\delta p_{\phi}-
c^2_{s\,\phi} \delta \rho_{\phi}\ , 
\end{equation}
where $c_{s \ , \phi}^2$ is the adiabatic sound speed of the scalar field, 
defined in equation (\ref{cs2}), which can also be written as 
\begin{equation}
\label{cs2bis}
c_{\phi}^2 = w_{\phi}-{1 \over 3 {\cal{H}} } {\dot{w}_{\phi} \over 1+w_{\phi}} \ .
\end{equation} 
In most Quintessence scenarios, the field is modeled as a 
component with negative, and slowly varying, equation of state, so that 
from eq. (\ref{cs2bis}), $c_{\phi}^2 \simeq w_{\phi}$. As 
discussed in \cite{GDM,GDM2}, looking at the continuity and Euler equations 
(\ref{continuity},\ref{euler}), a fluid with negative sound speed without 
entropy and stress terms would make adiabatic fluctuations unable to give a 
pressure support against gravitational collapse of density perturbations:
in other words, under these circumstances, the adiabatic pressure 
fluctuations would accelerate the collapse rather than oppose it, as can be 
derived from the density perturbations evolution in the limit of sub-horizon
scales. In this scenario, density perturbations would rapidly become non-linear
after entering the horizon, unless the entropic term in 
eq.(\ref{continuity}) acts as a stabilizing mechanism: this requires 
$w_{\phi} \Gamma_{\phi} > 0$. To check this possibility, W. Hu in 
\cite{GDM} introduces the effective sound speed $c_{eff\,\phi}^2$ 
defined in the rest frame of the scalar field, where 
$\delta T^0_{j\,\phi}=0$ (in \cite{GDM}, these considerations are applied to 
a more general ``generalized dark matter'' component, which can recover the 
Quintessence scalar field
case, as well as matter and radiation). Following 
this approach, we write the gauge invariant entropic term as 
\begin{equation}
\label{entropic}
w_{\phi} \Gamma_{\phi}=(c_{eff}^2-c_{\phi}^2) \delta_{\phi}^{(rest)}\ ,
\end{equation}
where $\delta_{\phi}^{(rest)}$ is the density contrast in the dark 
energy rest frame, which therefore coincides with the gauge 
invariant density perturbation $\Delta_{\phi}$ as follows: 
\begin{equation}
\label{deltarest}
\delta_{\phi}^{(rest)}=\Delta_{\phi}=
\delta_{\phi}+3{{\cal{H}} \over 
k}\left(1+w_{\phi}\right)\left(v_{\phi}-B\right)\ .
\end{equation}
By doing so, the stabilization mechanism of scalar field perturbations is 
expressed only in terms of gauge invariant quantities. In the 
mentioned case of $c_{s\,\phi}^{2} < 0$, effective pressure support  
is obtained if the entropic term (\ref{entropic}) is positive.
The effective sound speed can be interpreted as a rest frame sound speed; 
importantly,  it allows to define a stabilization scale for a 
perturbation, given by the corresponding effective sound horizon.  This 
formalism has been used in \cite{GDM} to show that  density perturbations 
in a minimally coupled scalar field of Quintessence are damped out below 
the horizon, so that the Quintessence rapidly becomes a smooth component: 
in this case, indeed, it can be verified that the effective sound speed is 
$c_{eff\,\phi}^2=1$, giving a relativistic behavior to the corresponding 
density fluctuations. 

The situation in Extended Quintessence scenarios can however be much different, 
in what the effective sound speed may be strongly affected by the presence 
of the non-minimal coupling term. Rewriting $c_{eff}^2$ as 
\begin{equation}
\label{ceff2}
c_{eff\,\phi}^2={\delta p_{\phi}+c_{s\,\phi}^{2}
\, 3{\cal{H}}h_{\phi}(v_{\phi}-B)/k
\over
\delta\rho_{\phi}+3{\cal H}h_{\phi}(v_{\phi}-B)/k}\ ,
\end{equation}
where $h_{\phi}\equiv\rho_{\phi}+p_{\phi}$, 
it is quite evident that, on sub-horizon scales $k\gg {\cal H}$, 
$c_{eff\,\phi}^{2}\simeq {\delta p_{\phi} / \delta \rho_{\phi}}$. As
discussed in the previous Section, in  $\delta \rho_{\phi}$ lies the main
difference between ordinary Quintessence and Extended Quintessence: while
${\delta p_{\phi}/\delta \rho_{\phi}} \simeq 1 $ for minimally coupled 
scalar fields, giving relativistic values to $c_{eff}^2$ and damping field
perturbations, this ratio may be much lower than unity whenever the energy
density perturbations of the scalar field are enhanced by perturbations in
the matter field, and this is a peculiar property of non-minimally coupled 
scalar fields. 

Another genuine feature which is expected in EQ scenarios regards 
the viscosity of the dark energy component. As pointed out, again 
in \cite{GDM}, the anisotropic stress can also be a 
smoothing mechanism for the scalar field, damping density perturbations 
through its effects on velocity perturbations in equation (\ref{euler}). 
A viscosity parameter $c_{vis}^2$ is introduced to relate 
velocity/metric shear and anisotropic stress; for the Quintessence scalar 
field, we have 
\begin{equation} 
\label{cvisc2}
\dot{\pi}_{T\,\phi} + 3 {\cal H} {\pi}_{T\,\phi}=
{4 c_{vis\,\phi}^2\over w_{\phi}}\left(kv_{\phi}-\dot{H}_{T}\right)\ .
\end{equation} 
In the limit of negligible $\dot{\pi}_{T\,\phi}$ and in shear-free frames 
($H_{T}=0$), $c_{vis}^2 > 0$ determines a viscous damping of velocity 
perturbations, as it can be seen through the Euler equation (\ref{euler}), 
which sums up with the viscosity effect arising from the 
cosmological expansion; 
thus, if $c_{vis}^2> 0$, viscosity can be an extra 
smoothing mechanism. 
On the other hand, $c_{vis}^{2}<0$ results in a term which 
acts against the cosmological viscosity into the Euler equation. 

The viscosity parameter turns out to be zero for minimally coupled 
Quintessence, where anisotropic stress is not present: in that case, 
however, $c_{eff}^{2}=1$, so that the adiabatic stress only is enough 
in smoothing the scalar field on sub-horizon scales. On the other 
hand, for Extended  Quintessence fields, we expect a non zero 
contribution to  $c_{vis}^2$, due to  the traceless part of 
$\delta T^{i\, nmc}_{j}[\phi]+\delta T^{i\, grav}_{j}[\phi]$. 

\begin{figure} 
\centerline{ 
\psfig{file=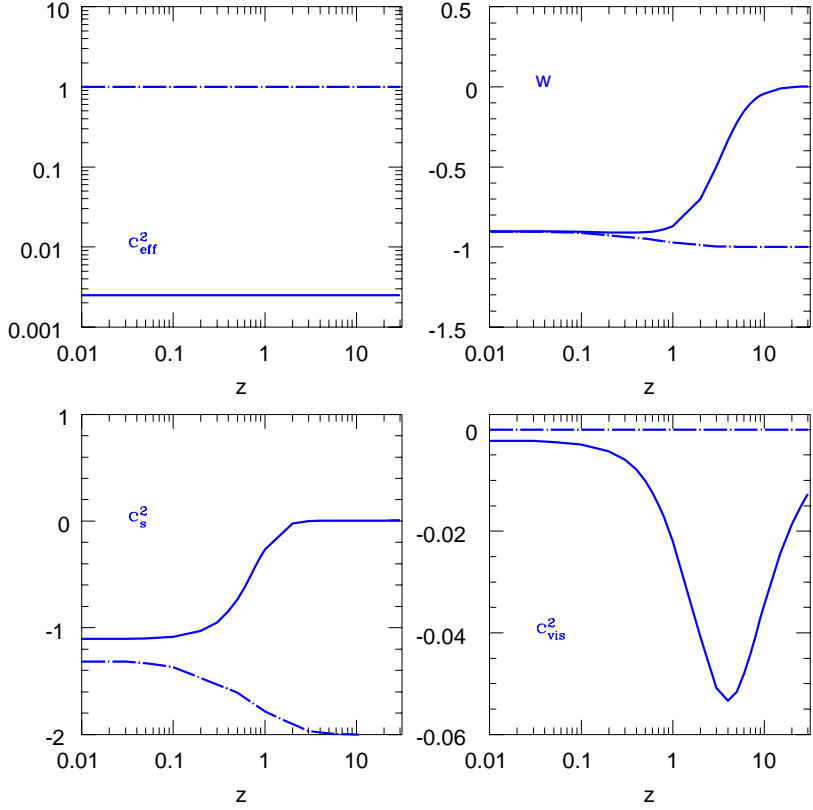,height=5.in,width=5.in}
} 
\caption{Redshift behavior of effective sound speed, equation of 
state, sound speed and viscosity in tracking EQ (solid) and Q 
(dotted dashed) scenarios.} 
\label{f3}
\end{figure}

We plot the four quantities $c_{eff\,\phi}^{2}$, $w_{\phi}$, 
$c_{s\,\phi}^{2}$, $c_{vis\,\phi}^{2}$ in figure \ref{f3}, 
comparing results in our tracking EQ scenario (solid lines) 
and in an ordinary minimally coupled Quintessence cosmology 
(dotted dashed).

\begin{figure} 
\centerline{ 
\psfig{file=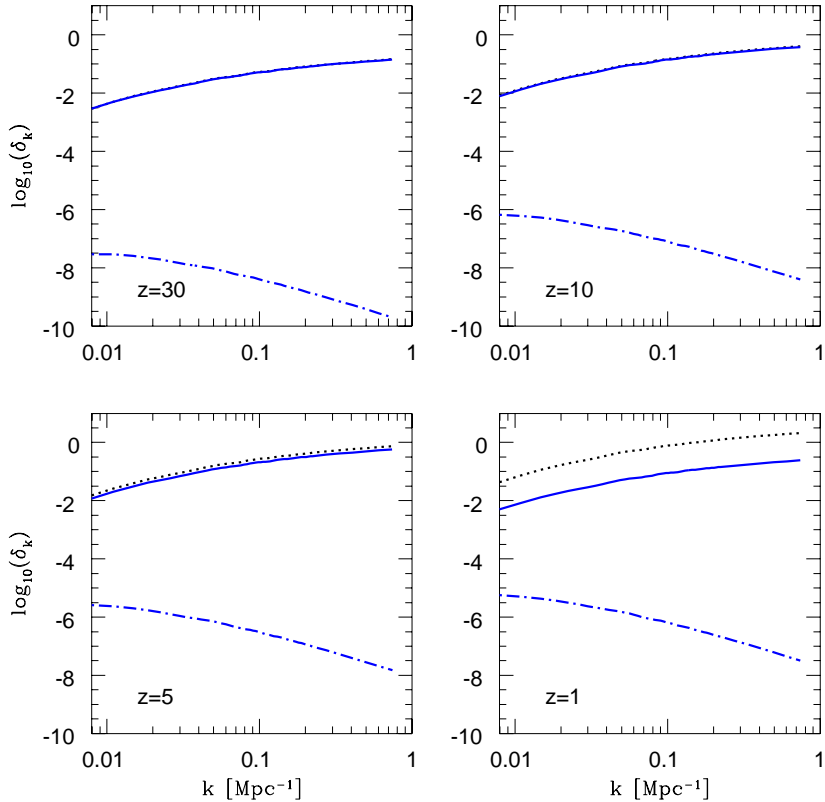,height=5.in,width=5.in} 
} 
\caption{Spectral power of density fluctuations for 
matter (dotted) and dark energy in tracking EQ (solid) and Q 
(dotted dashed) scenarios at different redshifts.} 
\label{f2}   
\end{figure}

The most striking effect is for $c_{eff\,\phi}^{2}$. 
For all the redshifts relevant for structure formation 
the effective dark energy sound speed is vanishing in the EQ 
case, allowing for a behavior of its density perturbations 
analogous to that of non-relativistic matter. Correspondingly, 
minimally coupled Quintessence has $c_{eff\,\phi}^{2}=1$. 
This reproduces the same result as in figure \ref{f2} obtained 
with a different approach. The more the gravitational term in 
$\delta\rho_{\phi}$ dominates in the denominator of the expression
(\ref{ceff2}), the larger is the suppression of the dark energy 
pressure reaction to the density contrast growth. 
Even if the plotted results are strictly valid only in our model, 
we stress that this is an example of a general occurrence 
in scalar-tensor dark energy cosmologies. In addition, 
since the gravitational term in (\ref{dT00grav}) is 
proportional to $(1/\kappa -F)$ which can in general 
assume both positive and negative values, 
even the sign of $c_{eff\,\phi}^{2}$ can be reversed 
realizing a scenario in which the sound speed 
accelerates the collapse on scales smaller than 
the sound horizon instead of opposing it. 

An analogous behavior can be found by looking at the plots of 
$w_{\phi}$ and $c_{s\,\phi}^{2}$, in what that they 
are severely depressed at relevant redshifts in EQ models. 
Also, the difference between $c_{s\,\phi}^{2}$ and $w_{\phi}$ is due 
to the time derivative of the equation of state through equation 
(\ref{cs2bis}). Notice in particular that in minimally coupled 
Quintessence we have $c_{s\,\phi}^{2},w_{\phi}<0$ while 
$c_{eff\,\phi}=1$; as we already stressed, the latter quantity is indeed the 
appropriate one to explain the behavior of a minimally coupled 
scalar field, resulting in the relativistic damping of sub-horizon
perturbations. 
Finally it is interesting to check whether viscosity can be 
sufficiently effective in damping out density and velocity perturbations, 
even when adiabatic pressure fluctuations are not. As in the 
minimally-coupled case, it turns out that this is not the case; 
first, the amplitude of $c_{vis\,\phi}^{2}$ is much lower than unity, 
and second the negative sign yield an enhancement of velocity 
perturbations, instead of a damping, as we stressed above; the peak 
at $z\simeq 1$ is due to the onset of cosmic acceleration, i.e. the 
sign change into the cosmic equation of state. 
 
In practice, there are no mechanisms to slow or decrease the 
amplitude of Extended Quintessence density perturbations in the 
gravitational dragging regime. 
This analysis confirms, and better clarifies, the results of previous 
Section concerning the scalar field density fluctuations power spectrum. 

\section{Concluding remarks}
\label{conclusions}
We studied the behavior of linear perturbations 
in scalar-tensor cosmologies, focusing on the density 
fluctuations of the scalar field $\phi$ coupled 
with the Ricci scalar $R$. 
We found that such coupling can activate a 
``gravitational dragging" 
of the scalar field density fluctuations by 
the cosmological metric perturbations, which in turn 
are powered by the whole cosmological stress-energy 
tensor through the Einstein equations. That is, 
as the non-minimal coupling represents 
a power exchange channel between the scalar field 
component and the general relativistic 
cosmological gravitational potentials, we studied in 
particular how such channel acts at the level of 
linear density perturbations in the scalar field, 
represented in particular by the density contrast 
$\delta_{\phi}=\delta\rho_{\phi}/\rho_{\phi}$. 
In conditions in which $\phi$ is not the dominant cosmological 
component, the power injection coming from Gravity can largely 
dominate $\delta_{\phi}$ forcing its dynamics to be similar to 
that of the dominant component itself. On the other hand, in the 
same conditions, the scalar field contributes to the 
cosmological gravitational potentials by a fraction 
given by the ratio between the scalar field and total energy 
densities. 

This phenomenology has important consequences on the 
dark energy clustering properties in Extended Quintessence 
scenarios, where the non-minimally coupled scalar field 
is assumed to be responsible for the cosmic acceleration today. 
Namely, the dark energy assumes the features of a pressureless fluid 
when non-relativistic matter {\it m} dominates, i.e. after matter 
radiation equality and in the pre-accelerating stage of the cosmic 
expansion. In other words, the scalar field density perturbations can 
grow on sub-horizon scales, tracing those in the matter component; 
this fact is depicted in the scalar field density contrast 
$\delta_{\phi}$, as well as in the properties of its effective sound 
speed $c_{eff\,\phi}$: 
\begin{equation}
\label{clustering}
\delta_{\phi}\simeq\delta_{m}\ ,\ c_{eff\,\phi}^{2}\ll 1\ .
\end{equation}
As we already mentioned, the reason of this behavior lies in the 
gravitational coupling to the Ricci scalar, contributing a gravitational 
term in the scalar field energy density which gets the dominant 
contribution from the perturbation in the matter component. 
The latter perturbations are therefore able to feed the 
dark energy density fluctuations up to a large amount even if the 
non-minimal coupling is small enough to respect all the existing 
constraints on scalar-tensor theories of Gravity. 
We stress also that the behavior (\ref{clustering}) is not 
depending on the particular form assumed to describe the 
non-minimal coupling; indeed, such gravitational dragging regime 
holds whenever the contributions due to the non-minimal coupling 
dominate both in $\rho_{\phi}$ and $\delta\rho_{\phi}$, so that 
their ratio is rather insensitive to the detailed shape of 
such coupling. Moreover, it should be noticed that the behavior 
(\ref{clustering}) is not to be related to variations 
of the expectation value of the scalar field $\phi$; indeed, 
our study shows that density perturbations of a non-minimally 
coupled scalar field are sourced both by fluctuations 
of expectation value as well and by perturbations of the Ricci tensor. 
In particular, the dragging effect emphasized here is generated even in 
the limiting case of a homogeneous scalar field, being induced by the 
coupling with $R$. 

We have provided a worked example of the above phenomenology in 
Extended Quintessence scenario, involving 
a quadratic coupling between the field and $R$. Numerical 
integrations of the cosmological equation system shows that 
the dynamical condition (\ref{clustering}) is satisfied at 
redshifts relevant for the structure formation process, respecting 
all the existing constraints on scalar-tensor Gravity theories. 

We believe that these results open new perspectives on the 
standard picture of structure formation in dark energy cosmologies, 
since the gravitational dragging expressed by (\ref{clustering}) 
implies that both dark energy and matter exit the linear regime 
on sub-horizon cosmological scales at relevant redshifts. 
This immediately poses the problem of their evolution afterwards, 
i.e. the gravitational clustering of large overdensities and 
deep cavities composed by matter and scalar energy tangled by a 
non-minimal gravitational interaction; 
while as we already stressed the gravitational dragging 
is rather insensitive to the detailed shape of the non-minimal 
coupling, the same could be untrue at a non-linear level. 
In particular, for a given model, it would be interesting to look 
at the appearance of the resulting density profile after 
virialization, since this aspect could be constrained by observed 
rotational curves in nearby galaxies. 

\acknowledgements
We are grateful to Sabino Matarrese and Massimo Pietroni for 
their precious suggestions; we warmly acknowledge Diego Torres for 
stimulating discussions.  We also thank Christof Wetterich and Paolo 
Salucci for helpful comments. 

\section{Appendix: scalar field perturbations in synchronous gauge }
\label{appendice}
Numerical integration and analytic treatment of the  perturbation
equations get simplified when developed in the synchronous gauge 
\cite{MB}. To obtain our numerical results, we used a modified 
version of {\tt cmbfast} \cite{SZ} which covers Extended Quintessence 
cosmologies. 

A scalar-type metric perturbation in the synchronous gauge is
parameterized as
\begin{equation}
ds^2=a^2 [-d\tau ^2 + (\delta_{i j }+h_{i j })dx^i dx^j ] \  \ ,
\end{equation}        
\begin{equation}
h_{i j}( {\bf x}, \tau ) = \int d^3 k e^{i {\bf k} \cdot {\bf x}}
\left[
{\bf {\hat{k}_i  \hat{k}_j}}  h( {\bf k}, \tau)
+ ( {\bf {\hat{k}_i  \hat{k}_j}}- {1 \over 3} \delta_{ i j} )
6 \eta  ( {\bf k}, \tau)\right]\ ,
\end{equation}
being $h$  the trace of $h_{ij}$. 
By choosing $A=B=0$ in (\ref{hmunu}), the metric perturbations 
$H_L$ and $H_T$ are related to $h$ and $\eta$ by the following relations:
\begin{equation}
\nonumber
H_T= - {h \over 2}-3 \eta\ ,\ H_L={h \over 6}\ . 
\end{equation}
Defining the shear perturbation $\sigma $ by the relation 
$\sigma= a^2/k^2 p\pi_T Y$
respectively, one has, for the scalar 
field, the following quantities derived from the conserved 
$T^{\mu}_{\nu}[\phi]$ (eq. \ref{Ttotal}), in synchronous gauge:
$$
\delta \rho =
\delta {\rho}_{fluid} + \omega {\dot{\phi} \delta\dot{\phi} \over 
a^2}
+{1 \over 2} \left({ \dot{\phi}^2 {\omega}_{,\phi} \over  a^2}
-{1\over\kappa}f_{,\phi}+2V_{,\phi}\right) \delta \phi
-3{ {\cal H} \delta \dot{F} \over a^2} -
\left(  -{ R \over 2 } + {k^2 \over a^2} \right) \delta F -
{\dot{F}\dot{h} \over 2a^{2}}-
$$
\begin{equation}
\label{deltarhosyn}
-3 { {\cal{H}}^2 \over a^2 } \delta F + 
\left({1\over\kappa}-F\right){2 \over a^2} 
\left[ - { {\cal{H}}\dot{h} \over 2}  + k^2 \eta \right]\ ,
\end{equation}
 
$$
\delta p =   \delta p_{fluid} +
 \omega {\dot{\phi} \delta\dot{\phi} \over a^2}
+ {1 \over 2} \left({ \dot{\phi}^2 {\omega}_{,\phi} \over  a^2}
+{1\over\kappa}f_{,\phi}-2V_{,\phi}\right) \delta \phi
+{\delta \ddot{F}\over a^{2}} + { {\cal H} \delta \dot{F} \over a^2} +
\left(  -{R  \over 2 } + {2k^2 \over 3a^2} \right) \delta F
+{1 \over 3 } {\dot {F} \dot{h}  \over a^2} + 
$$
\begin{equation}
\label{deltapsyn}
+{\delta F \over a^2}  (2 \dot{\cal{H}}+ {\cal{H}}^2) +
{2 \over 3 a^2}\left({1\over\kappa}-F\right) 
\left[ -{\cal{H}}\dot{h} - {\ddot{h} \over 2 } + k^2 \eta
\right]\ ,
\end{equation}                  

\begin{equation}
\label{thetasyb}
(p+ \rho) v = (p_{fluid}+ \rho_{fluid} ) v_{fluid} 
+{k^2 \over a^2 } \left[
  \omega \dot{\phi} \delta \phi +\delta\dot{F } -{\cal H} \delta F
+2\left({1\over \kappa}-F\right)\dot{\eta} \right]
\end{equation}           
 
$$
(p+ \rho) \sigma=   (p_{fluid}+ \rho_{fluid} ) \sigma_{fluid}+  
$$
\begin{equation}
\label{shearsyn}
+ {2k^2 \over 3a^2 }
 \left[    \delta F + 3 { \dot{F} \over k^2} \left( \dot{\eta} + {\dot{h}
\over 6} \right)+\left({1\over \kappa}-F\right)\left( -{{\cal{H}} \over k^2}
\dot{h}-{6{\cal{H}} \over k^2}\dot{\eta} - 
{2 \over k^2}\ddot{h} + {12 \over k^2}\ddot{\eta} +\eta
\right)  \right]  \ , 
\end{equation}      
where $\delta p_{\phi} \equiv p_{\phi} \pi_{L\,\phi}$ is the 
isotropic pressure  perturbation. 
The perturbed Klein-Gordon equation reads
\begin{equation}
\nonumber
\delta \ddot{\phi} + \left( 2 {\cal H} + {\omega_{,\phi} \over \omega}
\dot{\phi} \right)\delta\dot{\phi} + \left[ k^2 + {\left( {\omega_{,\phi}
\over \omega} \right) }_{,\phi} {\dot{\phi}^{2} \over 2 } +
a^2 {\left( {-f_{, \phi}/k +2V_{, \phi} \over 2 \omega } \right) 
}_{,\phi} \right] \delta \phi=  -{\dot{\phi} \dot{h} \over 2 } + {a^2 
\over 2 \omega}  {f_{, \phi  R} \over k } \delta R \ .
\end{equation}
These perturbations enter in the perturbed Einstein equations, easy to 
solve in this gauge: 
\begin{eqnarray}                                
\label{t00}
k^{2}\eta -{1\over 2}{\cal H}\dot{h} &=& -{a^2\kappa \delta \rho
\over 2}  \ ,\\
\label{ti0}
k^{2}\dot{\eta} &=& { a^{2}\kappa (p+\rho)v \over 2}  \ ,\\
\label{tii}
\ddot{h}+2{\cal H}\dot{h}-2k^{2}\eta &=& -3 a^{2}\kappa \delta p \
,\\
\label{tij}          
 \ddot{h}+6\ddot{\eta}+2{\cal 
H}(\dot{h}+6\dot{\eta})-2k^{2}\eta &=&
-3 a^{2}\kappa (p+\rho)\sigma\ .
\end{eqnarray}
This set of differential equations requires initial conditions on the 
metric and fluid perturbations; we adopt adiabatic initial conditions 
\cite{PB}).


\begin{thebibliography}{}

\bibitem{PERL} S. Perlmutter et at., Astrophys.J. 517, 565 (1999)

\bibitem{RIESS} A. Riess et al., Astron.J. 116, 1009 (1998);
 A. Riess et al., Astrophys. J. 536, 62 (2000)

\bibitem{TURNER} M.S. Turner and A. Riess, submitted to Astrophys.J,
astro-ph/0106051 (2001); A.G. Riess et al., Astrophys.J.560, 49 (2001)

\bibitem{CMBDATA} C. B. Netterfield et al., submitted to Astrophys. J., 
preprint astro-ph/0104460 (2001); P. De Bernardis et al., submitted to 
Astrophys. J., astro-ph/0105296 (2001); A. T. Lee et al., submitted to 
Astrophys. J. Lett., astro-ph/0104459 (2001); R. Stompor et al., 
submitted to Astrophys. J. Lett., astro-ph/0105062 (2001); N. W. 
Halverson et al., submitted to Astrophys. J., astro-ph/0104489 (2001); 
C. Pryke et al., submitted to Astrophys. J., astro-ph/0104490 (2001)

\bibitem{LSS} J.A. Peacock et al., Nature 410, 169 (2001)

\bibitem{CCP} S.M. Carroll, Living Rev.Rel. 4, 1 (2001)

\bibitem{RP} B. Ratra, P.J. Peebles, Phys. Rev. D 37, 3406 (1988)

\bibitem{W} C. Wetterich, Nucl. Phys. B 302, 645 (1988) 

\bibitem{TRACK}  P. J. Steinhardt, L. Wang, I. Zlatev, Phys.Rev. D59, 
123504 (1999) 


\bibitem{LS} A.R. Liddle, R.J. Scherrer, Phys.Rev. D59, 023509 (1999) 

\bibitem{K0} T. Chiba, T. Okabe, M. Yamaguchi, Phys.Rev. D62, 023511 (2000)

\bibitem{K} C. Armendariz-Picon, V. Mukhanov, P.J. Steinhardt, 
Phys.Rev. D63, 103510, (2001)

\bibitem{EQ} F. Perrotta, C. Baccigalupi, S. Matarrese, Phys.Rev. D61  
023507 (2000)

\bibitem{UZANNMC} J. P. Uzan, Phys.Rev. D59 123510 (1999)

\bibitem{AMENMC} L. Amendola, Phys.Rev. D60 043501 (1999) 

\bibitem{CHIBANMC} T. Chiba, Phys.Rev.  D60, 083508 (1999) 

\bibitem{BP} N. Bartolo, M. Pietroni, Phys.Rev. D61, 023518 (2000)

\bibitem{RU} A. Riazuelo and J.P. Uzan, Phys.Rev. D62, 083506 (2000)

\bibitem{CHIBACOS} T. Chiba, Phys.Rev. D64, 103503 (2001)

\bibitem{TEQ} C. Baccigalupi, S. Matarrese, F. Perrotta, Phys.Rev. D62  
123510 (2000)

\bibitem{BCMB} C. Baccigalupi, A. Balbi, S. Matarrese, F. Perrotta, N.
Vittorio, Phys.Rev.D in press (2002)

\bibitem{CDM} D.H. Lyth, A.Riotto, Phys.Rept. 314, 1 (1999)
 
\bibitem{WETTECLUMPS} C. Wetterich, Phys.Lett. B 522, 5 (2001) 

\bibitem{MATOS}  F. S. Guzman, T. Matos, Class.Quant.Grav. 17, L9 (2000) 

\bibitem{MATOS2} T. Matos, F. S. Guzman, D. Nunez, Phys.Rev. D62  061301 
(2000) 

\bibitem{ALS} A. Arbey, J. Lesgourgues, P. Salati, Phys.Rev. D64  
123528 (2001)

\bibitem{LH} E.L. Lokas, Y. Hoffman, astro-ph/0108283, submitted to MNRAS

\bibitem{GDM} W. Hu, ApJ 506, 485 (1998)

\bibitem{KS} H. Kodama, M. Sasaki, Prog.Theor. Phys. Suppl., 78, 1 (1984)

\bibitem{HW} J.C. Hwang, ApJ 375, 443 (1991); 
J.C. Hwang, Class.Quantum Grav. 7, 1613 (1990)

\bibitem{EFP} G. Esposito-Farese, D. Polarski, Phys.Rev. D63, 063504 (2001)

\bibitem{Faraoni} V. Faraoni, Phys.Rev. D62  023504 (2000) 

\bibitem{BD} N.D.Birrel and P.C.W.Davies, {\it Quantum fields in curved 
spacetime}, Cambridge University Press 1982

\bibitem{GDM2} W. Hu, D. Eisenstein,  Phys.Rev.D 59, 083509 (1999) 

\bibitem{Bardeen} J. M. Bardeen, Phys. Rev. D 22, 1882 (1980) 

\bibitem{JBD} J. O. Dickey et al., Science 265, 482 (1994); J. G.  
Williams,  X. X. Newhall, J. O. Dickey, Phys. Rev. D 53, 6730 (1996); T.  
M. Eubkans et al., Bull. Am. Phys. Soc., Abstract \# K 11.05 (1997) 

\bibitem{PB} F. Perrotta, C. Baccigalupi, Phys.Rev. D59 123508 (1999)

\bibitem{NUCLEO} X. Chen, R.J. Scherrer, G. Steigman, 
Phys.Rev. D63, 123504 (2001)

\bibitem{BMR} P.Brax, J.Martin, A.Riazuelo, Phys.Rev.D62, 103505 (2000)

\bibitem{MB} C.P. Ma, E. Bertschinger, ApJ 455, 7 (1995). 

\bibitem{SZ} U. Seljak, M. Zaldarriaga, Astrophys.J 469, 437, (1996)

\end{thebibliography}
\end{document}